\documentclass[twocolumn,journal]{IEEEtran}
\usepackage[T1]{fontenc}
\usepackage[latin9]{inputenc}
\usepackage{amsmath}
\usepackage{amssymb}
\usepackage{graphicx}
\usepackage[unicode=true,
 bookmarks=true,bookmarksnumbered=true,bookmarksopen=true,bookmarksopenlevel=1,
 breaklinks=false,pdfborder={0 0 0},pdfborderstyle={},backref=false,colorlinks=false]
 {hyperref}
\hypersetup{pdftitle={Your Title},
 pdfauthor={Your Name},
 pdfpagelayout=OneColumn, pdfnewwindow=true, pdfstartview=XYZ, plainpages=false}

\usepackage{fancyhdr}

\makeatletter

\newcommand{\lyxmathsym}[1]{\ifmmode\begingroup\def\b@ld{bold}
  \text{\ifx\math@version\b@ld\bfseries\fi#1}\endgroup\else#1\fi}

\let\oldforeign@language\foreign@language
\DeclareRobustCommand{\foreign@language}[1]{%
  \lowercase{\oldforeign@language{#1}}}

\usepackage[caption=false,font=footnotesize]{subfig}

\makeatother

\begin{document}
\title{Highly Sensitive Coupled Oscillator Based on an Exceptional Point
of Degeneracy and Nonlinearity}
\author{Alireza Nikzamir, Filippo Capolino \thanks{A. Nikzamir and F. Capolino are with the Department of Electrical
Engineering and Computer Science, University of California, Irvine,
CA 92697, USA. e-mail: \protect\href{http://anikzami,\%20f.capolino@uci.edu}{anikzami, f.capolino@uci.edu}.}}
\markboth{Journal}{Your Name \MakeLowercase{\emph{et al.}}: Department of Electrical
Engineering and Computer Science, University of California, Irvine,
CA 92697, USA}
\maketitle

\thispagestyle{fancy}

\begin{abstract}
We propose a scheme for obtaining highly-sensitive oscillators in
a coupled-resonator system with an exceptional point of degeneracy
(EPD) and a small instability. The oscillator with the exceptional
degeneracy is realized by using two coupled resonators with an almost
balanced small-signal gain and loss, that saturates due to nonlinear
effects of the active component, resulting in an oscillation frequency
that is very sensitive to a perturbation of the circuit. Two cases
are investigated, with two parallel LC resonators with balanced small-signal
gain and loss that are either coupled wirelessly by mutual inductance
or coupled-wired by a capacitor. This paper demonstrates theoretically
and experimentally the conditions to obtain a second-order EPD oscillator
and analyzes the ultrasensitivity of the oscillation frequency to
components' perturbation, including the case of asymmetric perturbation
that breaks PT-symmetry. We discuss the effects of nonlinearity on
the performance of the oscillator and how the proposed scheme improves
the sensing's sensitivity of perturbations. In contrast to previous
methods, our proposed degenerate oscillator can sense positive or
negative changes of a circuit component. The degenerate oscillator
circuit may find applications in various areas such as ultrasensitive
sensors, tunable oscillators and modulators.
\end{abstract}


\IEEEpeerreviewmaketitle{}

\section{Introduction}

\IEEEPARstart{O}{scillators} are fundamental components of radio
frequency (RF) electronics. Traditionally, an oscillator is viewed
as a positive feedback mechanism utilizing a gain device with a selective
reactive circuit. An oscillator generates a continuous, periodic single-frequency
output when the Barkhausen\textquoteright s criteria are satisfied.
The oscillator circuit should have a self-sustaining mechanism such
that noise gets filtered, quickly grows and becomes a periodic signal.
Most RF oscillators are implemented by only one active device for
noise and cost considerations, such as Van der Pol and voltage-controlled
oscillators \cite{Vanderpol1934Thenonlinear}. Oscillators can be
realized by a simple LC resonator with positive feedback using a negative
resistance. Pierce, Colpitts, and tunnel diode oscillators play a
role of negative resistance in a circuit, as well as a cross-coupled
transistor pair \cite{Razavi1998RF,Pierce1923Piezoelectric,Colpitts1927Oscillation}.
All oscillators are based on a single-pole operation, i.e., a single
pole is rendered unstable when the system is brought above the threshold.
While oscillators based on an LC resonator are the most common type
of oscillator, other designs may feature distributed \cite{Wu2001Silicon,Tanaka2002Synchronizability},
ring \cite{Hajimiri1999Jitter,Nayak2017Low}, coupled \cite{Chang1997Phase},
or multi-mode \cite{Endo1976Mode} oscillators, which come with their
own challenges and advantages.

In this paper, we study the concept of an oscillator based on a double
pole, i.e., an oscillator designed to utilize an exceptional point
of degeneracy (EPD) in two coupled resonators. A system reaches the
EPD when at least two eigenmodes coalesce into a single degenerate
one, in their eigenfrequencies (eigenvalues) and polarization states
(eigenvectors) \cite{Heiss1990Avoided,Vishik_1960_Solution,Seyranian1993Sensitivity,Lancaster1964On,Kato1966Perturbation,Bender2002Generalized,Heiss2012Thephysics,Peng2014Parity}.
The letter ``D'' in EPD refers to the key concept of ``degeneracy''
where the relevant eigenmodes, including the associated eigenvectors
are fully degenerate \cite{Berry2004Physics}. The degeneracy order
refers to the number of coalescing eigenfrequencies. The concept of
EPD has been implemented traditionally in systems that evolve in time,
like in coupled resonators \cite{Sakhdari2017PT-symmetric,Schindler2011Experimental,Nikzamir2021Demonstration,Stehmann2004Observation,Rouhi2022Exceptional},
periodic and uniform multimode waveguides \cite{Figotin2005Gigantic,Othman2017Theory,Othman2015Demonstration,Sloan2018Theory,Abdelshafy2019Exceptional,Mealy2020General},
and also in waveguides using Parity-Time (PT) symmetry \cite{Bender1998Real,Klaiman2008Visualization,Abdelshafy2019Exceptional}.
EPDs have been recently demonstrated also in temporally-periodic single
resonator without a gain element \cite{Kazemi2019Exceptional,kazemi2019experimental,Kazemi2021Ultra},
inspired by the finding that EPD exists in spatially periodic lossless
waveguides \cite{Nada2017Theory,Othman2016Giant,Figotin2003Oblique},
using the non-diagonalizability of the transfer matrix associated
to the periodic system.

A very significant feature of a system with EPD is the ultra-sensitivity
of its eigenvectors and eigenvalues to a perturbation of a system's
parameter. This property paves the way to conceive a scheme to measure
a small change in either physical, chemical, or biological parameters
that causes a perturbation in the system. Typically, a sensor's sensitivity
is related to the amount of spectral shift of a resonance mechanism
in response to a perturbation in environmental parameters, for example,
a glucose concentration or other physical variations like changing
pressure, etc. Sensors with EPD can be wired or wirelessly connected
to the measuring part of the sensor. In this paper, we show the extreme
sensitivity of an oscillator operating at an EPD to external perturbations.

Previous parity-time (PT)-symmetric circuits have been conceived as
two coupled resonators where changes happen at one resonator, and
the data is detected on the other side \cite{Chen2018Generalized}.
When the circuit is perturbed away from its EPD, PT-symmetry must
be maintained in order to obtain two real-valued frequencies. For
example, in Ref. \cite{Chen2018Generalized}, when one side's capacitance
is perturbed, the authors tuned the other side's capacitance using
a varactor to keep the PT-symmetry in the circuit, so they can still
observe two real-valued shifted frequencies perturbed away from the
degenerate EPD frequency. Thus, in previously published schemes (implementing
the demonstration of sensitive measurement of a perturbation) the
exact value of such perturbation should be exactly known to tune the
other side of the system in order to keep the circuit PT-symmetric.
This seems to contradict the idea that the circuit is used as a sensor
of an unknown measurable quantity. That scheme could be saved if combined
with an iterative method performing an automatic scan to reconstruct
the PT-symmetry. Anyway, this rebalancing procedure (to keep the system
PT-symmetric) makes it more complicated to use of such a scheme when
designing a sensor.

The limitation of PT-symmetry schemes is that they can detect only
perturbations that lead to the same-sign change in a system's component,
such as a capacitor's value. This is because a PT-symmetric system
provides two real-valued frequencies only when the system is perturbed
away from its EPD in one direction (for example for\textit{ $G$}
values smaller than the \textit{$G_{e}$} related to the EPD, when
looking at the eigenfrequencies in Fig. \ref{fig:Circuit}). If the
perturbation makes the system move in the other direction, the shift
of the frequencies is in the imaginary parts \cite{Heiss2004Exceptional,Schindler2011Experimental,Chen2018Generalized,Stehmann2004Observation},
leading to two complex-valued frequencies and hence to instability.
One must also consider that any mismatch between the sensor side (typically
the part with losses) and the reader side (typically the part with
gain), even involuntary, leads to an asymmetric system. Thus, a PT-symmetric
system in practice always shows two complex-valued eigenfrequencies
and increase the risks of self-sustained oscillations (unless an EPD
is designed having a large enough damping factor, larger than the
eigenfrequency perturbation due to circuit tolerances). Noise and
nonlinearities play a critical role in the robustness of these kinds
of applications and affect the possibility of instability \cite{Wiersig2020Robustness}.
Some error-correction techniques are studied in \cite{Kananian2020Coupling}
to overcome some of these drawbacks using a nonlinear PT-symmetry
scheme to enhance the robustness of sensing.

In this paper, we provide a scheme that starts by using a quasi PT-symmetric
condition, working near an EPD, that makes the double-pole system
slightly unstable even before having any perturbation. In other words,
we turn the above-mentioned practical problems that occur in PT-symmetric
systems to our advantage when the circuit has to be used in a highly
sensitive sensor. We set the gain value slightly higher than the loss
counterpart to make the system slightly unstable. As a result of instability
and nonlinear gain, the signal grows until the active gain component
reaches saturation, and the working operation will be close to the
EPD.

We first show the behavior of wirelessly coupled LC resonators through
the dispersion relation of the resonance frequency versus perturbation
and we discuss the occurrence of EPDs in such a system. In section
\ref{sec:Oscillator-charactaristics}, we use the nonlinear model
for the gain to achieve the oscillator\textquoteright s characteristics.
We show that the oscillation frequency is very close to the EPD frequency.
The EPD-based oscillator has an oscillation frequency that is very
sensitive to perturbation, exhibiting the typical square root-like
behavior of EPD systems, where the change in frequency of the oscillator
is proportional to the square root of the perturbation. In section
\ref{sec:Sensor-point-of}, we demonstrate the highly sensitive behavior
of the circuit by breaking PT-symmetry, i.e., by perturbing the capacitance
on the lossy side (the sensing capacitance). In this case, the circuit
oscillates at a shifted frequency compared to the EPD one. Notably,
both positive and negative perturbations in the capacitance are shown
to lead to opposite shifted frequencies, i.e., the proposed scheme
detects positive and negative changes in the capacitance, in contrast
to conventional PT-symmetry systems \cite{Schindler2011Experimental,Chen2018Generalized,Sakhdari2017PT-symmetric}
that generate frequency shifts associated to only one sign of the
perturbation. The EPD is demonstrated also by analyzing the bifurcation
of the dispersion diagram at the EPD frequency by using the Puiseux
fractional power series expansion \cite{Welters2011On_Explicit,Kato1966Perturbation}.
In section \ref{sec:Wired-Coupling}, we show the condition to have
an EPD in two resonators coupled by a capacitor and demonstrate the
occurrence of the EPD by using the Puiseux series and experimentally
using a nonlinear active element. Also, we discuss how noise contributes
to the system by showing the power spectrum of the system and the
phase noise. The theoretical results are in a good agreement with
the experimental results, pointing out that small perturbations in
the system can be detected by easily measurable resonance frequency
shifts, even in the presence of thermal noise and electronic noise.
The advantages of using the proposed circuit as an ultra-sensitive
sensor and how the experimental results show that the oscillator is
sensitive to both positive and negative capacitance changes are discussed
in Section \ref{sec:Wired-Coupling}. Very sensitive sensors based
on the oscillator scheme discussed here can be a crucial part of various
medical, industrial, automotive and aerospace applications that require
sensing physical or chemical changes as well as biological quantities.

\section{Oscillator based on coupled resonators with EPD\label{sec:Coupled-Oscilator}}

\begin{figure}[t]
\begin{centering}
\includegraphics[width=0.85\columnwidth]{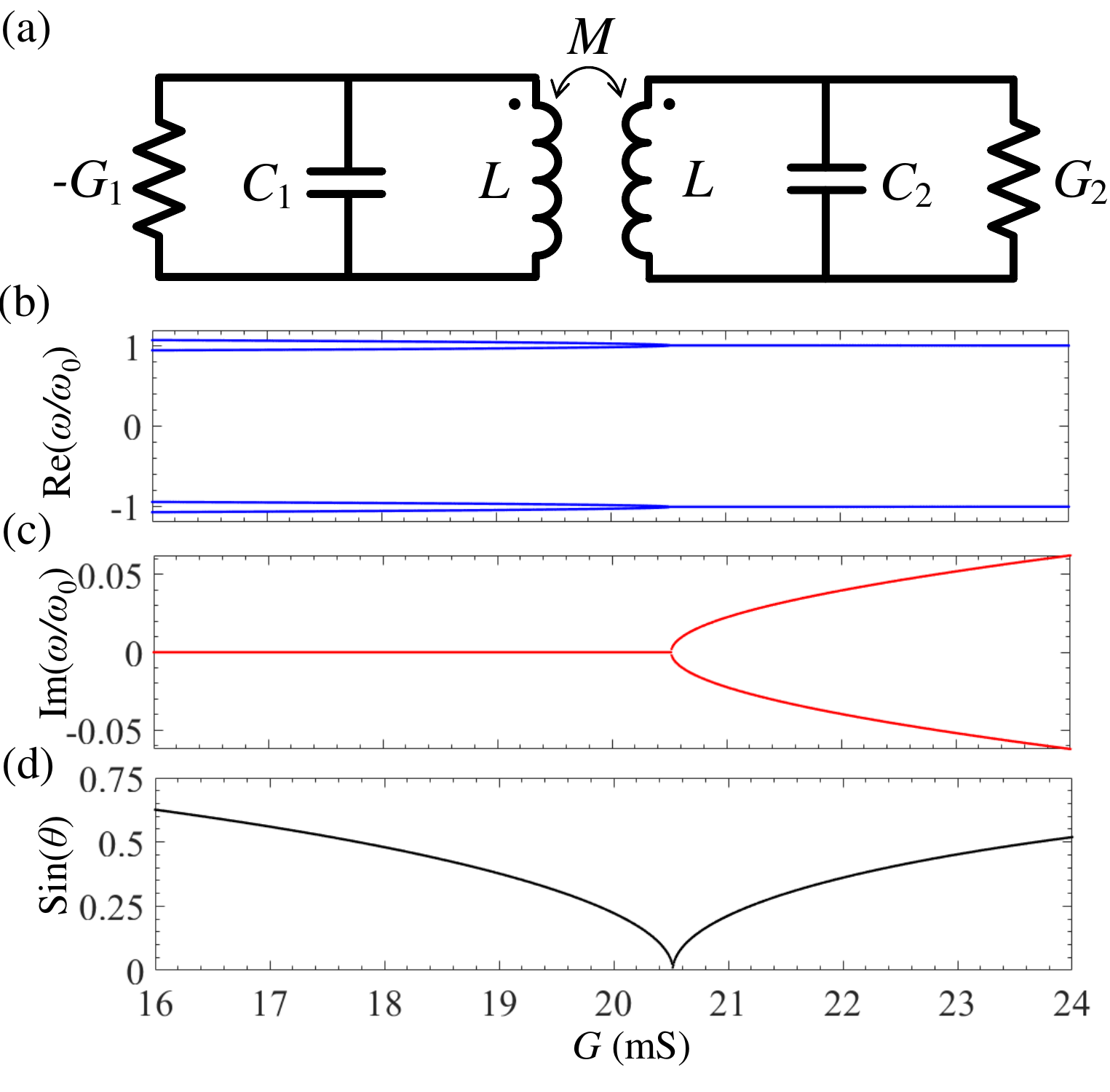}
\par\end{centering}
\centering{}\caption{\label{fig:Circuit}(a) Coupled resonators terminated with linear
$-G_{1}$ on the gain side ($n=1$) and $G_{2}$ on the loss side
($n=2$), with $G_{1}=G_{2}=G$, and inductances $L=0.1\:\mathrm{\mu H}$,
mutual coupling $k=M/L=0.2$, capacitances of $C_{n}=C_{0}=1\:\mathrm{nF}$
($n=1,2$). The natural frequency of each (uncoupled) LC resonator
is $\omega_{0}=1/\sqrt{LC_{0}}=10^{8}\:\mathrm{s}^{-1}$. Normalized
eigenfrequencies of the coupled circuit are calculated by using Eqs.
(\ref{eq:eigenfrequency1}) and (\ref{eq:eigenfrequency2}). (b) Real,
and (c) imaginary parts of the resonance angular frequencies normalized
by $\omega_{0}$ varying $G$ on both sides of the EPD value. (d)
At the EPD point ($G=G_{e}=20.52\,\mathrm{mS}$, $\omega_{e}=1.01\times10^{8}\:\mathrm{s}^{-1}$),
two state eigenvectors coalesce demonstrated by the vanishing of $\mathrm{sin}(\theta)$.}
\end{figure}

We investigate the coupled resonators shown in Fig. \ref{fig:Circuit}(a),
where one parallel LC resonator is connected to gain (left side, or
$n=1$) and the other is connected to loss (right side, or $n=2$).
In this ideal circuit, negative conductance (gain) has the same magnitude
as the loss to exactly satisfy PT-symmetry. When a system satisfies
PT-symmetry, it means that the system is invariant to the application
of the two operators \textquotedbl P\textquotedbl{} and \textquotedbl T\textquotedbl .
The \textquotedbl P\textquotedbl{} stands for parity transformation
(making a spatial reflection (e.g., $x\rightarrow-x$)), and \textquotedbl T\textquotedbl{}
stands for time-reversal transformation ($t\rightarrow-t$), where
$x$ is the coordinate and $t$ is the time.

By writing Kirchhoff's current laws, we obtain the equations

\begin{equation}
\begin{cases}
\frac{d^{2}Q_{1}}{dt^{2}}=-\frac{1}{LC_{1}\left(1-k^{2}\right)}Q_{1}+\frac{k}{LC_{2}\left(1-k^{2}\right)}Q_{2}+\frac{G_{1}}{C_{1}}\frac{dQ_{1}}{dt}\\
\frac{d^{2}Q_{2}}{dt^{2}}=+\frac{k}{LC_{1}\left(1-k^{2}\right)}Q_{1}-\frac{1}{LC_{2}\left(1-k^{2}\right)}Q_{2}-\frac{G_{2}}{C_{2}}\frac{dQ_{2}}{dt}
\end{cases}\label{eq:KVL}
\end{equation}
where $Q_{n}$ is the capacitors charge on the gain side ($n=1$)
and the lossy side ($n=2$), and $\dot{Q}_{n}=dQ_{n}/dt$ is the current
flowing into the capacitor. We define the state vector as $\boldsymbol{\Psi}(t)\equiv[Q_{1},Q_{2},\dot{Q}_{1},\dot{Q}_{2}]^{\mathrm{T}}$,
consisting of a combination of stored charges and currents on both
sides, and the superscript $\mathrm{T}$ denotes the transpose operation.
Thus, we describe the system in a Liouvillian formalism as

\begin{equation}
\begin{array}{c}
\frac{d\boldsymbol{\Psi}}{dt}=\underline{\boldsymbol{\mathrm{M}}}\boldsymbol{\Psi},\\
\\
\underline{\boldsymbol{\mathrm{M}}}=\left(\begin{array}{cccc}
0 & 0 & 1 & 0\\
0 & 0 & 0 & 1\\
-\frac{1}{LC_{1}\left(1-k^{2}\right)} & \frac{k}{LC_{2}\left(1-k^{2}\right)} & \frac{G_{1}}{C_{1}} & 0\\
\frac{k}{LC_{1}\left(1-k^{2}\right)} & -\frac{1}{LC_{2}\left(1-k^{2}\right)} & 0 & -\frac{G_{2}}{C_{2}}
\end{array}\right).
\end{array}\label{eq:Liouvillian}
\end{equation}

We are interested in finding the eigenfrequencies and eigenvectors
of the system matrix $\underline{\boldsymbol{\mathit{\mathrm{M}}}}$
describing the circuit. Assuming signals of the form $Q_{n}\varpropto e^{j\omega t}$,
we write the eigenvalues problem associated with the circuit equations,
$(\underline{\boldsymbol{\mathit{\mathrm{M}}}}-j\omega\underline{\boldsymbol{\mathrm{I}}})\boldsymbol{\Psi}=0$,
where $\underline{\boldsymbol{\mathrm{I}}}$ is a $4$ by $4$ identity
matrix. Then, by solving $P(\omega)\triangleq\mathrm{det}(\underline{\boldsymbol{\mathit{\mathrm{M}}}}-j\omega\underline{\boldsymbol{\mathrm{I}}})=0$,
the four eigenfrequencies are found. By assuming $C_{1}=C_{2}=C_{0}$
and linear $G_{1}=G_{2}=G$, a symmetry condition that has been described
as PT symmetric \cite{Stehmann2004Observation}, the characteristic
equation takes the simplified form
\begin{equation}
\begin{array}{c}
P(\omega)=\left(1-k^{2}\right)\left(\frac{\omega}{\omega_{0}}\right)^{4}\:\:\:\:\:\:\:\:\:\:\\
\\
\:\:\:\:+\left(G^{2}Z^{2}\left(1-k^{2}\right)-2\right)\left(\frac{\omega}{\omega_{0}}\right)^{2}+1=0,
\end{array}\label{eq:char EQ1}
\end{equation}
where $Z=\sqrt{L/C_{0}}$ is a convenient normalizing impedance, and
$\omega_{0}^{2}=1/\left(LC_{0}\right)$. The characteristic equation
is quadratic in $\omega^{2}$; therefore, $\omega$ and $-\omega$
are both solutions. Moreover, the $\omega$'s coefficients in the
characteristic equation are real, hence $\omega$ and $\omega^{*}$
are both solutions, where {*} represents the complex conjugate operation.
The 4 by 4 matrix $\underline{\boldsymbol{\mathrm{M}}}$ results in
4 angular eigenfrequencies which are found analytically as,

\begin{equation}
\omega_{1,3}=\pm\omega_{0}\sqrt{\frac{1}{1-k^{2}}-\frac{G^{2}Z^{2}}{2}-\sqrt{b}},\label{eq:eigenfrequency1}
\end{equation}

\begin{equation}
\omega_{2,4}=\pm\omega_{0}\sqrt{\frac{1}{1-k^{2}}-\frac{G^{2}Z^{2}}{2}+\sqrt{b}},\label{eq:eigenfrequency2}
\end{equation}

\begin{equation}
b=-\frac{1}{1-k^{2}}+\left(\frac{G^{2}Z^{2}}{2}-\frac{1}{1-k^{2}}\right)^{2}.\label{eq:EPD cond.}
\end{equation}

The EPD frequency is found when the component values obey the condition

\begin{equation}
b=0.\label{eq:EPD condition}
\end{equation}

So far $b=0$ is a necessary condition, but in a simple system like
this, the eigenvectors can be represented as a function of the eigenvalues
so this condition is also sufficient to show the convergence of the
eigenvectors, hence for an EPD to occur. Under this condition, we
calculate the EPD angular frequency based on Eqs. \ref{eq:eigenfrequency1}
and \ref{eq:EPD condition} as $\omega_{1}=\omega_{2}=\omega_{e}$
where

\begin{equation}
\omega_{e}=\frac{\omega_{0}}{\sqrt[4]{1-k^{2}}}.\label{eq:EPD Frequency}
\end{equation}

The real and imaginary parts of the eigenfrequencies are shown in
Fig. \eqref{fig:Circuit}(b) and (c) varying $G$. It is seen from
this plot that the system's eigenfrequencies are coalescing at a specific
balanced linear gain/loss value $G=G_{e}$, where $b=0$. Note that
in this scenario, the EPD-enabling value $G_{e}$ is derived from
Eq. \eqref{eq:EPD condition} as

\begin{equation}
G_{e}=\frac{1}{Z}\left(\frac{1}{\sqrt{1-k}}-\frac{1}{\sqrt{1+k}}\right).\label{eq:GEPD}
\end{equation}

For clarification, when $G=0$ (lossless and gainless circuit), we
have two pairs of resonance frequencies $\omega_{1,3}=\pm\omega_{0}/\sqrt{1+k}$
and $\omega_{2,4}=\pm\omega_{0}/\sqrt{1-k}$, and $\omega_{1}\neq\omega_{2}$
always, except for the trivial case with $k=0$, when these eigenfrequencies
are equal to those of the isolated circuits, but since the two circuits
are isolated this is not an important degeneracy. With the given values
of $L$ and $C$ in the caption of Fig. \ref{fig:Circuit}, a second-order
EPD occurs when $G=G_{e}=20.52\,\mathrm{mS}$. In this case, the circuit's
currents and charges grow linearly with increasing time as $Q_{n}\varpropto t\cos(\omega_{e}t)$,
and they oscillate at the degenerate frequency $\omega_{e}$. Also,
near the EPD point, the eigenfrequencies, when perturbing \textit{$G$},
have a square root-like behavior as $|\omega-\omega_{e}|\propto\pm\sqrt{\left(GZ\right)^{2}-\left(G_{e}Z\right)^{2}}$\cite{Schindler2011Experimental}.
A second coalescence (i.e., degeneracy) happens for larger values
of \textit{$G$}, i.e., at $G_{e}^{'}=\frac{1}{Z}\left(\frac{1}{\sqrt{1-k}}+\frac{1}{\sqrt{1+k}}\right)$.
When $G>G_{e}^{'}$ all frequencies are imaginary, so we only study
cases of $G<G_{e}^{'}$. In the strong coupling regime, $0<G<G_{e}$,
the eigenfrequencies are purely real, and the oscillation wave has
two fundamental frequencies. In the weak coupling regime, $G_{e}<G<G_{e}^{'}$,
the frequencies are complex conjugate and the imaginary part of the
angular eigenfrequencies is non-zero, and it causes two system solutions
($Q_{1}$ and $Q_{2}$) with damping and exponentially growing signals
in the system. Since the solution of the circuit is $Q_{n}\varpropto e^{j\omega t}$,
the eigenfrequency with a negative imaginary part is associated to
an exponentially growing signal and the oscillation frequency is associated
to the real part of the eigenfrequency.

At each positive (real part) angular eigenfrequency $\omega_{1}$
and $\omega_{2}$, calculated by Eqs. \eqref{eq:eigenfrequency1}
and \eqref{eq:eigenfrequency2}, we find the two associated eigenvectors
$\boldsymbol{\Psi}_{\boldsymbol{1}}$ and $\boldsymbol{\Psi_{2}}$
by using Eq. \eqref{eq:Liouvillian}. A sufficient condition for an
EPD to occur is that at least two eigenvectors coalesce, and that
is what we check in the following. Various choices could be made to
measure the state vectors' coalescence at an EPD, and here, the Hermitian
angle between the state amplitude vectors $\boldsymbol{\Psi_{1}}$
and $\boldsymbol{\Psi_{2}}$ is defined as

\begin{equation}
\theta=\mathrm{arccos}\left(\frac{|<\boldsymbol{\Psi_{1}},\boldsymbol{\Psi_{2}}>|}{||\boldsymbol{\Psi_{1}}||\:||\boldsymbol{\Psi_{2}}||}\right).\label{eq:theta}
\end{equation}

Here the inner product is defined as $<\boldsymbol{\Psi_{1}},\boldsymbol{\Psi_{2}}>=\boldsymbol{\Psi_{1}^{\dagger}}\boldsymbol{\Psi_{2}}$,
where the dagger symbol $\dagger$ denotes the complex conjugate transpose
operation, | | represents the absolute value, and || || represents
the norm of a vector. According to this definition, the state vectors
$\boldsymbol{\Psi_{1}}$ and $\boldsymbol{\Psi_{2}}$ correspond to
resonance frequencies $\omega_{1}$ and $\omega_{2}$, respectively.
When some system's parameter is varied, eigenfrequencies and associated
eigenvectors are calculated using Eq. \eqref{eq:Liouvillian}. In
the case when $G$ varies, Fig. \eqref{fig:Circuit}(d) shows that
the sine of the angle $\theta$ between the two eigenvectors vanishes
when the eigenfrequencies coalesce, which indicates the coalescence
of the two eigenmodes in their eigenvalues and eigenvectors and hence
the occurrence of a second-order EPD.

\begin{figure}[t]
\centering{}\includegraphics[width=0.95\columnwidth]{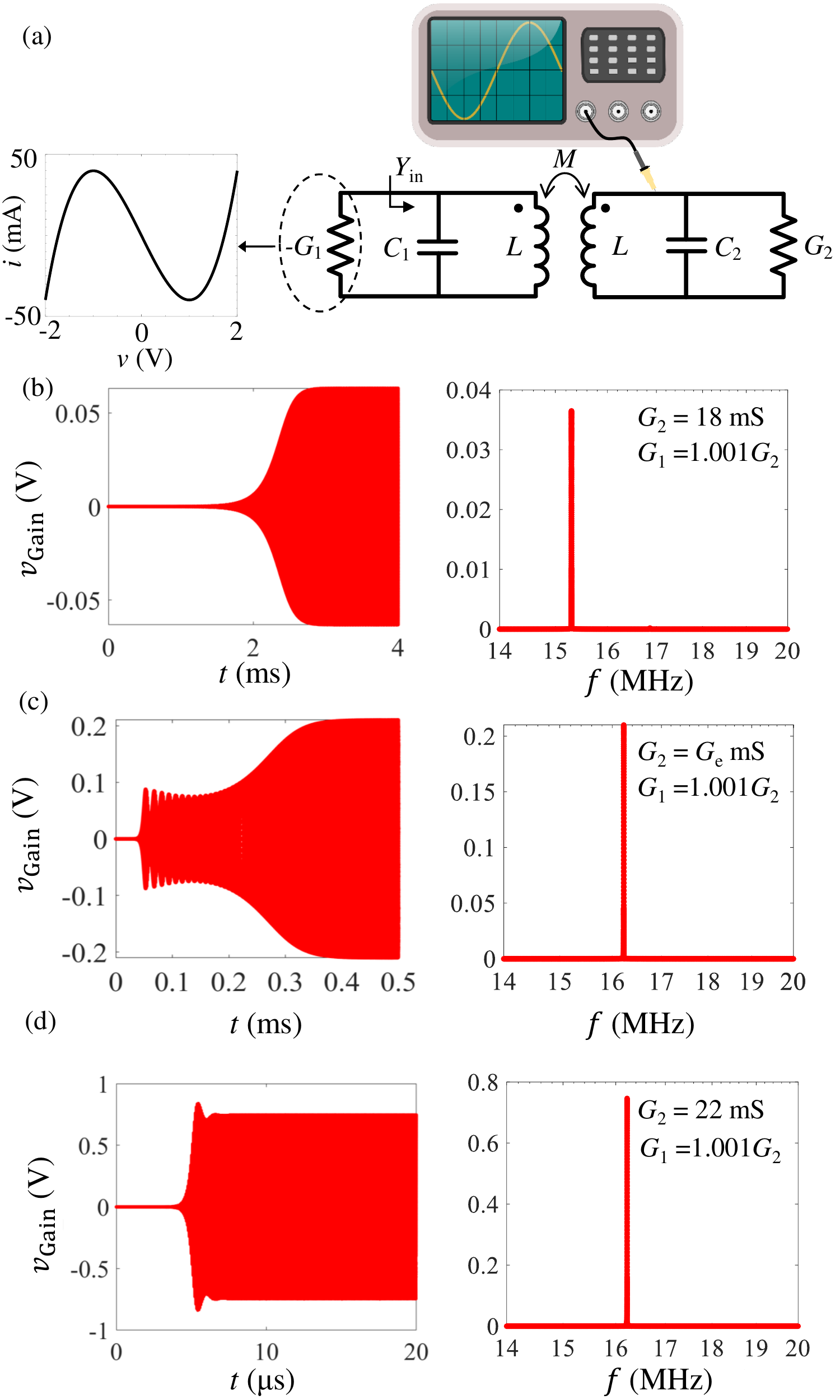}\caption{\label{fig:cubic transient}(a) Cubic gain $i-v$ curve with parameters
$G_{1}=G_{e}=20.52\,\mathrm{mS}$ and $\alpha=6.84\,\mathrm{mS}/\mathrm{V}^{2}$
(it corresponds to $V_{\mathrm{b}}=1\:\mathrm{V}$). Time-domain response
and frequency spectrum of the oscillatory signal with a cubic model
where the gain is always 0.1\% more than the loss (i.e., $G_{1}=1.001G_{2}$)
with: (b) $G_{2}\lesssim G_{1}<G_{e}$, (c) $G_{1}=1.001G_{e}$ and
$G_{2}=G_{e}$, and (d) $G_{1}\gtrsim G_{2}>G_{e}$, where $G_{e}=20.52\,\mathrm{mS}$.}
\end{figure}

\section{Oscillator characteristics\label{sec:Oscillator-charactaristics}}

This section describes the important features of an oscillator made
of two coupled resonators with discrete (lumped) elements with balanced
gain and loss, coupled wirelessly by a mutual inductance as in Fig.
\ref{fig:Circuit}. The transient time-domain, frequency spectrum,
and double pole (or zero, depending on what we look at) features are
discussed. A cubic model (nonlinear) of the active component providing
gain is considered. The parameters used here are the same as those
used in the previous section, where $G_{e}=20.52\,\mathrm{mS}$ leads
to an EPD of order two at a frequency of 16.1 MHz, except that $-G_{1}$
accounts also for the nonlinear part responsible for the saturation
effect.

\textbf{A. }Transient and frequency behavior

Time and frequency-domain responses of the coupled resonators circuit
are obtained by using the Keysight Advanced Design System (ADS) circuit
time-domain simulator, as shown in Fig. \ref{fig:cubic transient}(b)-(d).
The cubic model for gain, in Fig. \ref{fig:cubic transient}(a), represented
as

\begin{equation}
i=-G_{1}v+\alpha v^{3}\label{eq:cubic}
\end{equation}
is a simplified description of the gain obtained from a cross-coupled
transistor or an operational amplifier (opamp) based circuit. Here,
$-G_{1}$ is the small-signal gain provided by the negative slope
of the $i-v$ curve, i.e., is the negative conductance in the small-signal
region and $\alpha=G_{1}/\left(3V_{\mathrm{b}}^{2}\right)$ is a third-order
nonlinearity that describes saturation, where $V_{\mathrm{b}}$ is
a turning point voltage determined by the biasing direct current (DC)
voltage. We assume $V_{\mathrm{b}}=1\:\mathrm{V}$, and to start self-sustained
oscillation we assume that gain $-G_{1}$ is not a perfect balance
of the loss $G_{2}$. Indeed, we assume that $G_{1}$ is $0.1\%$
larger than $G_{2}$. Therefore, the system is slightly perturbed
away from the PT-symmetry condition to start with. We also assume
white noise (at the temperature of 298 K) is present in the loss resistor
and it is indeed the initial condition for starting oscillations.

Using $G_{1}$ to be $0.1$\% larger than $G_{2}$, the circuit is
unstable and it starts to oscillate, and after a transient, the circuit
saturates, yielding a stable oscillation, as shown in Fig. \ref{fig:cubic transient}(b)-(d).
As it was shown in Figs. \ref{fig:Circuit}(b) and (c) assuming linear
gain, for values of $G_{1}=G_{2}<G_{e}$, the system has two distinct
eigenfrequencies $\omega_{1}$ and $\omega_{2}$ with zero imaginary
part. However, when using the cubic nonlinear model with $G_{1}$=$1.001G_{2}$,
with $G_{2}\lesssim G_{1}<G_{e}$, the imaginary part is not zero
anymore because of the nonlinearity and slightly broken PT-symmetry.
Thus, when using the cubic model, after an initial transient, the
oscillation signal associated to the eigenfrequency with a negative
imaginary part dominates and makes the system saturates. Considering
again the inital result in Figs. \ref{fig:Circuit}(b) and (c) assuming
linear gain, it is noted that when $G_{1}=G_{2}>G_{e}$, we have two
complex conjugate eigenfrequencies, and the one associated to the
negative imaginary part makes the circuit oscillate. However, when
using the cubic gain model with $G_{1}$=$1.001G_{2}$, with $G_{1}\gtrsim G_{2}>G_{e}$,
eigenfrequencies approximately follow the linear gain eigenfrequency
trend. It means that for the values $G_{1}\gtrsim G_{2}>G_{e}$, we
have a larger negative imaginary part of the eigenfrequency than when
$G_{2}\lesssim G_{1}\leq G_{e}$. The rising time is related to the
magnitude of the negative imaginary part of the eigenfrequency; indeed,
as shown in Fig. \ref{fig:cubic transient}(b)-(d), the rising time
is different in the three cases. By going further from the EPD point,
the signal saturates in a shorter time. In all cases, the frequency
spectrum of the time-domain signal is found by taking the Fourier
transform of the voltage on the gain side after reaching saturation,
for a time window of $10^{3}$ periods.

\textbf{B. }Root locus of zeros of the total admittance

This subsection discusses the frequency (phasor) approach to better
understand the degenerate resonance frequencies of the coupled resonators
circuit. We use the admittance resonance method and we demonstrate
the occurrence of double zeros at the EPD. The resonance condition
based on the vanishing of the total admittance implies that

{\small{}
\begin{equation}
Y_{\mathrm{in}}(\omega)-G_{1}=\frac{P(\omega)}{j\frac{L}{\omega_{0}^{2}}\left(1-k^{2}\right)\omega^{3}+L^{2}G_{1}\left(1-k^{2}\right)\omega^{2}-jL\omega}=0,\label{eq:Roots}
\end{equation}
}where the $Y_{\mathrm{in}}$ is the input admittance of the linear
circuit, including the capacitor $C_{1}$, looking right as shown
in Fig. \ref{fig:cubic transient}(a). Here, we assume linear gain
with $G_{1}=G_{2}=G$, i.e., satisfying PT symmetry.

\begin{figure}[t]
\begin{centering}
\includegraphics[width=0.9\columnwidth]{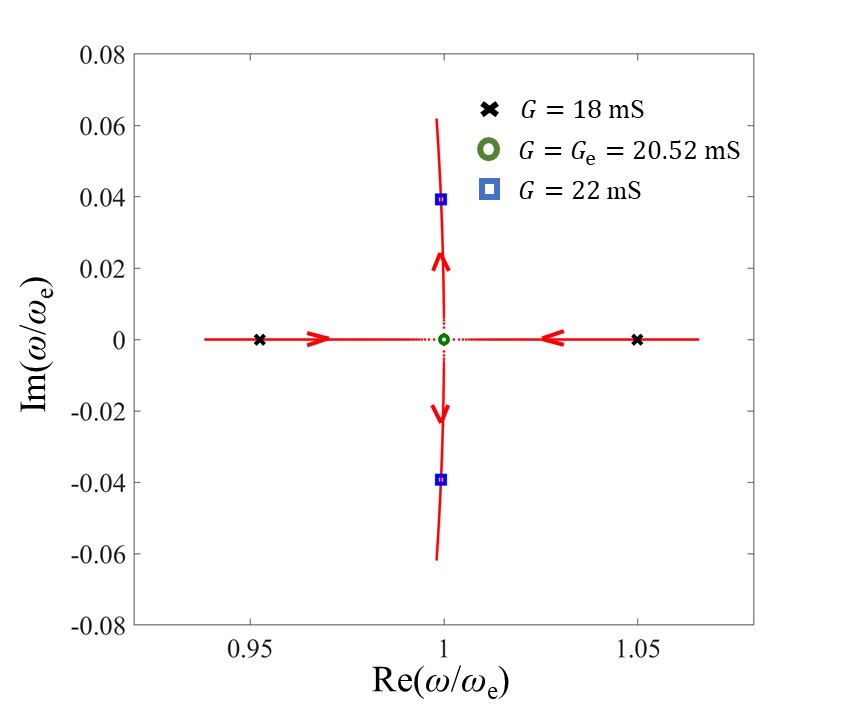}
\par\end{centering}
\centering{}\caption{\label{fig:Trajectory-of-zeros}The trajectory of the zeros of $Y_{in}(\omega)-G=0$
shows the two resonance frequencies by varying $G$ from $15\,mS$
to $25\,mS$ (we assume linear gain with $G_{1}=G_{2}=G$). When $G=G_{e}$,
the two branches meet at $\omega_{e}$. Therefore, at the EPD, the
frequency $\omega_{e}$ is a double zero of $Y_{in}(\omega)-G=0$.}
\end{figure}

The polynomial $\mathrm{P}(\omega)$ is given in Eq. \eqref{eq:char EQ1}.
We calculate the eigenfrequencies by finding the zeros of $Y_{\mathrm{in}}(\omega)-G$,
and this leads to the same $\omega$-zeros of $P(\omega)={\normalcolor \mathrm{det}}(\underline{\boldsymbol{\mathit{\mathrm{M}}}}-j\omega\underline{\boldsymbol{\mathrm{I}}})=0.$
Note that both $\omega(G)$ and $-\omega(G)$ are both solutions of
Eq. (\ref{eq:Roots}), as well as both $\omega(G)$ and $\omega^{*}(G)$.
The trajectories of the zeros of this equation, i.e., the resonance
frequencies $\omega(G)$, are shown in Fig. \ref{fig:Trajectory-of-zeros}
by varying linear $G$ from $18\,\mathrm{mS}$ to $G=22\,\mathrm{mS}$
(we recall that in this case $G=G_{1}=G_{2}$), in the complex frequency
plane. We show only the roots with $\mathrm{Re}(\omega)>0$ for simplicity.
At the EPD, $G=G_{e}=20.52\,\mathrm{mS}$, and the above equation
reduces to $Y_{\mathrm{in}}(\omega)-G\propto(\omega-\omega_{e})^{2}$,
i.e., the admittance exhibits a double zero at the EPD angular frequency
$\omega_{e}$. This unique property is also responsible for the square
root-like behavior of resonance frequency variation due to the perturbation
in a system, as discussed next, which is the key to high sensitivity.
Moreover, for values $G<G_{e}$, the two resonance frequencies are
purely real, and for $G>G_{e}$, the two resonance frequencies are
a complex conjugate pair.

\section{Sensor point of view\label{sec:Sensor-point-of}}

\textbf{A. High sensitivity and the Puiseux fractional power expansion}

As mentioned in the Introduction, when the system is operating at
an EPD, the eigenfrequencies are extremely sensitive to system perturbations,
and this property is intrinsically related to the Puiseux series \cite{Welters2011On_Explicit}
that provides a fractional power series expansion of the eigenvalues
in the vicinity of the EPD point. We consider a small perturbation
$\Delta_{\mathrm{X}}$ of a system parameter $X$ as
\begin{figure}[t]
\begin{centering}
\includegraphics[width=3.5in]{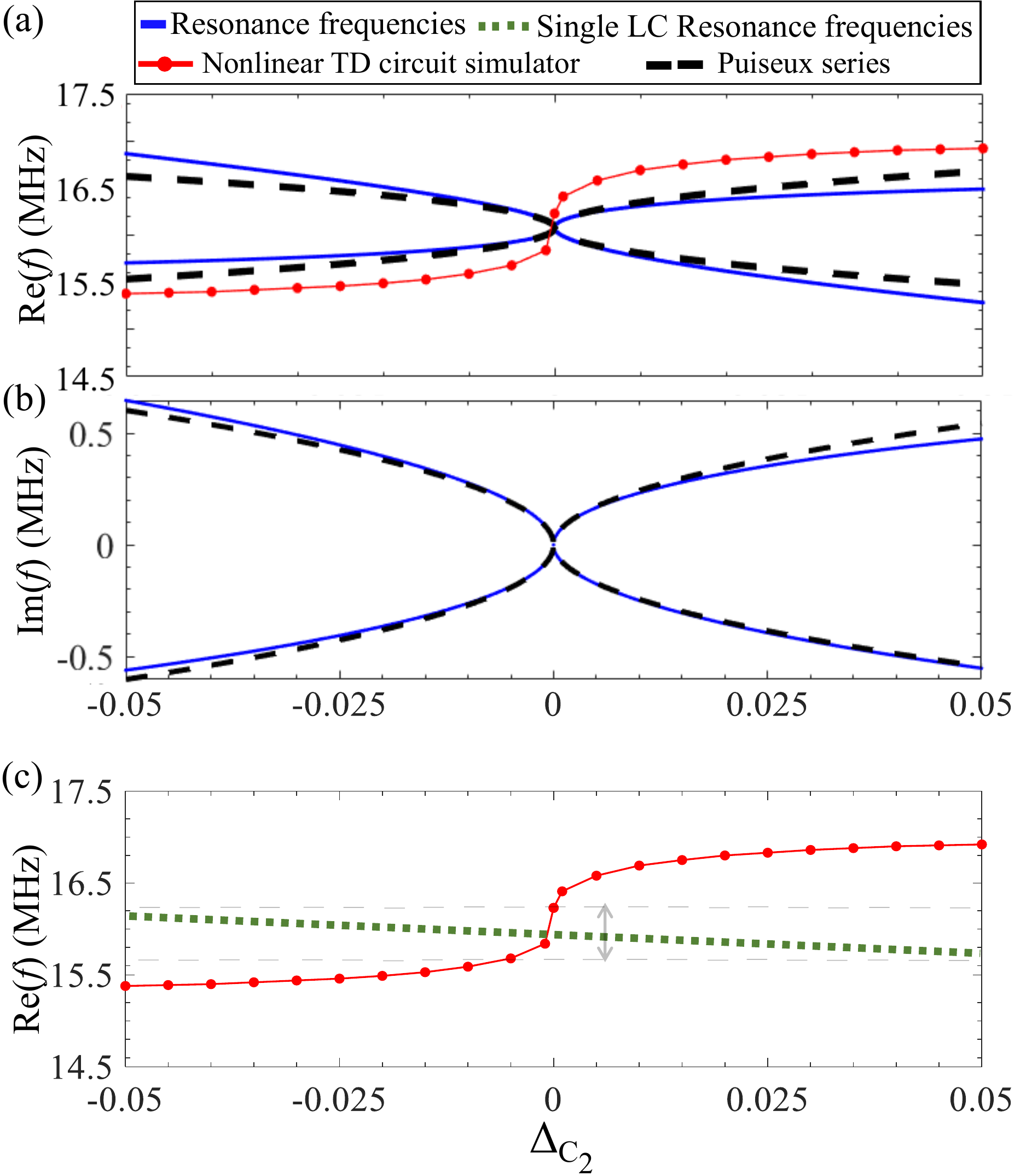}
\par\end{centering}
\caption{\label{fig:The-Puiseux-fractional} High sensitivity of the circuit
to a variation of capacitance $C_{2}$ . We show the (a) real and
(b) imaginary parts of the resonance frequencies (using linear gain)
when varying $C_{2}$, compared to the frequency of oscillation after
saturation when using nonlinear gain. Solid blue lines show the resonance
frequencies obtained by solving the characteristic equation Eq. \eqref{eq:char EQ1};
dashed lines show the estimate obtained by using the Puiseux fractional
power series expansion truncated to its first order. In both cases,
gain is a linear negative conductance with $G_{1}=G_{2}=G_{e}$. Red
dots in (a) show the oscillation frequencies using nonlinear gain;
results are obtained by using the time-domain circuit simulator Keysight
ADS using the small-signal negative conductance $-G_{1}$ with $G_{1}=1.001G_{e}$,
i.e., it has been increased by $0.1\%$ from its loss balanced value
$G_{e}$ (we recall that $G_{2}=G_{e}$). The frequencies of oscillation
are obtained by applying a Fourier transform of the capacitor $C_{1}$
voltage after the system reaches saturation, for each considered value
of $C_{2}$. (c) Sensitivity comparison with single linear LC resonator,
when varying $\Delta_{\mathrm{C_{2}}}$. The much higher sensitivity
of the EPD oscillator with double pole is clear. Note that the whole
frequency variation relative to the full perturbation range of capacitance
($-5\%<\Delta_{\mathrm{C_{2}}}<5\%)$ for the single LC resonator
could be achieved by only $1/10$ of the perturbation ($-0.5\%<\Delta_{\mathrm{C_{2}}}<0.5\%$)
when the EPD based circuit is used. The highest sensitivity of the
EPD circuit is shown for very small perturbations $\Delta_{\mathrm{C_{2}}}$.}
\end{figure}

\begin{equation}
\Delta_{\mathrm{X}}=\frac{X-X_{e}}{X_{e}},\label{eq: perturbation_touchstone}
\end{equation}
where $X$ is the perturbated value of a component, and $X_{e}$ is
the unperturbed value that provides the EPD of second order. A perturbation
$\Delta_{\mathrm{X}}$ leads to a perturbed matrix $\underline{\boldsymbol{\mathrm{M}}}(\Delta_{\mathrm{X}})$
and, as a consequence, it leads to two distinct perturbed eigenfrequencies
$\omega_{p}(\Delta_{\mathrm{X}})$, with $p=1,2$, near the EPD eigenfrequency
$\omega_{e}$ as predicted by the Puiseux series containing power
terms of $\Delta_{\mathrm{X}}^{\frac{1}{2}}$. A good approximation
of the two $\omega_{p}(\Delta_{\mathrm{X}})$, with $p=1,2$, is given
by the first order expansion
\begin{equation}
\omega_{p}(\Delta_{\mathrm{X}})\simeq\omega_{e}+(-1)^{p}\alpha_{1}\sqrt{\Delta_{\mathrm{X}}}.\label{eq:Puisex}
\end{equation}

Following \cite{Welters2011On_Explicit,Kato1966Perturbation}, we
calculate $\alpha_{1}$ as

\begin{equation}
\alpha_{1}=\sqrt{-\frac{\frac{\partial H(\Delta_{\mathrm{X}},\omega)}{\partial\Delta_{\mathrm{X}}}}{\frac{1}{2!}\frac{\partial^{2}H(\Delta_{\mathrm{X}},\omega)}{\partial\omega^{2}}}},\label{eq:alpha1}
\end{equation}
where $H(\varDelta,\omega)=\mathrm{det}[\underline{\boldsymbol{\mathit{\mathrm{M}}}}(\Delta)-j\omega\underline{\boldsymbol{\mathrm{I}}}]$,
and its derivatives are evaluated at the EPD, i.e., at $\Delta_{\mathrm{X}}=0$
and $\omega=\omega_{e}$.

\begin{figure}[t]
\begin{centering}
\includegraphics[width=3.5in]{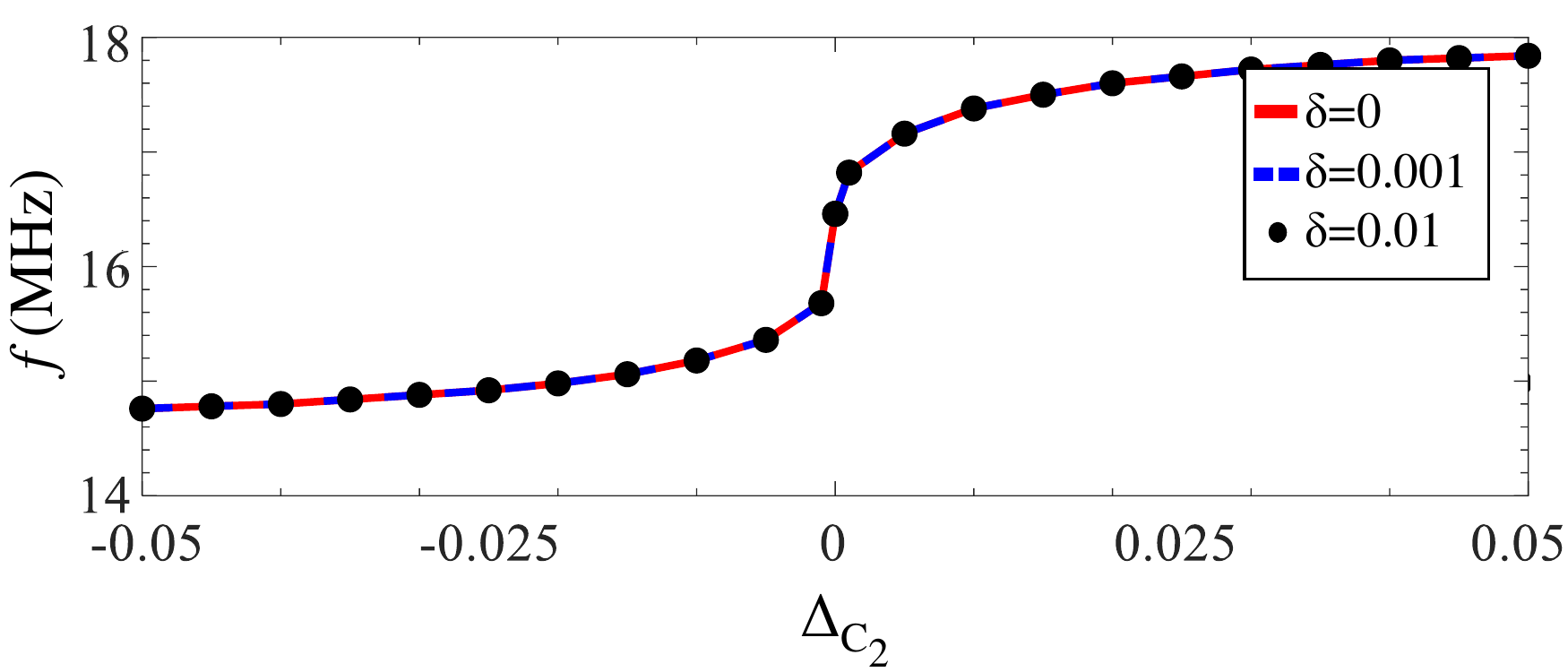}
\par\end{centering}
\caption{\label{fig:The-Puiseux-fractional-1} Robustness of the high sensitivity
of the circuit to a variation of capacitance $C_{2}$ .The oscillator's
fundamental frequencies of the circuit after each $0.5\%$ perturbation
on $C_{2}$ by using nonlinear gain are shown here, considering three
values of gain $G_{1}=G_{e}(1+\delta)$, where $G_{2}=G_{e}$, for
three different values of $\delta=0,\;0.001$, and $0.01$. These
three plots are on top of each other, meaning that even with a 1\%
mismatch between gain and loss, the oscillator's fundamental frequencies
are the same as those for smaller unbalanced situations. It is important
to note that both positive and negative perturbations of $C_{2}$
are detected.}
\end{figure}

Consider a coupled LC resonator, as described in Fig. \ref{fig:cubic transient}(a),
assume the capacitor $C_{2}$ on the loss side is perturbed from the
initial value as $(1+\Delta_{\mathrm{C_{2}}})C_{e}$, where $C_{e}$
is unperturbed value for both $C_{1}$ and $C_{2}$: the coefficient
$\alpha_{1}$ is found analytically as

{\small{}
\begin{equation}
\alpha_{1}=\sqrt{\frac{L^{2}\omega_{e}^{2}G_{e}^{2}\left(1+\frac{C_{e}\omega_{e}}{G_{e}}\right)\left(1-k^{2}\right)+\left(1-C_{e}L\omega_{e}^{2}\right)}{L^{2}\left(6C_{e}^{2}\omega_{e}^{2}+G_{e}^{2}\right)\left(1-k^{2}\right)-2C_{e}L}}.\label{eq:alpha1_1}
\end{equation}
}{\small\par}

The Puiseux fractional power series expansion Eq. \eqref{eq:Puisex}
indicates that for a small perturbation such that $|\Delta_{\mathrm{X}}|\ll1$,
the eigenfrequencies change dramatically from their original degenerate
value due to the square root function. The Puiseux series first-order
coefficient is evaluated by Eq. \eqref{eq:alpha1_1} as $\alpha_{1}=10^{7}(1.693+j1.530)\:\mathrm{rad/s}$.
The coefficient $\alpha_{1}$ is a complex number implying that the
system always has two complex eigenfrequencies, for any $C_{2}$ value.
In Fig. \eqref{fig:The-Puiseux-fractional} (a) and (b), the estimate
of $\omega_{p}$, with $p=1,2$, using the Puiseux series is shown
by a dashed black line. The calculated eigenfrequencies by directly
solving the characteristic equation Eq. (\ref{eq:char EQ1}) are shown
by solid blue lines. In this example, we can consider $C_{2}$ as
a sensing capacitance to detect possible variations in chemical or
physical parameters, transformed into electrical parameters, like
the frequency of oscillation in the circuit. For a small value of
$\Delta_{\mathrm{C_{2}}}$, around the EPD value $\Delta_{\mathrm{C_{2}}}=0$,
the imaginary and real parts of the eigenfrequencies experience a
sharp change, resulting in a very large shift in the oscillation frequency.
Note that this rapid change in the oscillation frequency is valid
for both positive and negative changes of $\Delta_{\mathrm{C_{2}}}$,
which can be useful for various sensing applications. Note also that
a perturbation of PT symmetry leads to instability.

To show how the sensitivity is improved when using the second-order
EPD (double-pole) oscillator, we compare its sensitivity to an analogous
scheme made of one single LC resonator, with an inductance of $L=0.1\:\mathrm{\mu H}$
and capacitance of $C_{2}=1\:\mathrm{nF}$ without adding gain or
loss. The resonance frequency of the LC resonator is $f_{0}=1/(2\pi\sqrt{LC_{2}})$
and by perturbing the capacitance $C_{2}$, the resonance frequency
changes as $f\thickapprox f_{0}(1-\Delta_{\mathrm{C_{2}}}/2)$. Figure
\eqref{fig:The-Puiseux-fractional} (c) shows the comparison between
two cases: (i) oscillation frequency of the EPD based oscillator with
nonlinear gain (red dots)using the time-domain circuit simulator Keysight
ADS, and (ii) the resonance frequency of the single LC resonator (dashed
green). The results demonstrate that the EPD-based circuit with nonlinearity
has higher sensitivity (square root-like behavior due to the perturbation)
than a single LC resonator without EPD (linear behavior). The whole
frequency variation, relative to the full perturbation range of capacitance
($-5\%<\Delta_{\mathrm{C_{2}}}<5\%)$ for the single LC resonator,
could be achieved by only $1/10$ of the perturbation ($-0.5\%<\Delta_{\mathrm{C_{2}}}<0.5\%$)
when the EPD based circuit is used. The highest sensitivity of the
EPD circuit is shown for very small perturbations $\Delta_{\mathrm{C_{2}}},$e.g.,
$\left|\Delta_{\mathrm{C_{2}}}\right|\approx1\%$. For larger $\Delta_{\mathrm{C_{2}}}$
variations, i.e., around $\left|\Delta_{\mathrm{C_{2}}}\right|\approx5\%$,
the slope of the flattened square root-like curve is similar to the
slope of the curve relative to the perturbed LC resonator.

To show how a telemetric sensor with nonlinearity works, we now consider
that the gain element is nonlinear, following the cubic model in Eq.
(\ref{eq:cubic}) where the small-signal negative conductance is $-G_{1},$
with value $G_{1}=1.001G_{e}$, i.e., increased by $0.1\%$ from its
loss balanced value $G_{e}$ as discussed earlier to make the circuit
slightly unstable and start self oscillations. The capacitor $C_{2}$
on the lossy side is perturbed by $\pm0.5\%$ steps and we perform
time-domain simulations using the circuit simulator implemented in
the Keysight ADS circuit simulator. Noise is assumed in the lossy
element $G_{2}$ to start oscillations. The time-domain voltage signal
at the capacitor $C_{1}$ on the gain side is read, and then, we take
the Fourier transform of such signal, after reaching saturation, for
a time window of $10^{3}$ periods. The oscillation frequency evolution
by changing $\Delta_{\mathrm{C_{2}}}$ is shown in Fig. \ref{fig:The-Puiseux-fractional}
by red dots. There is no imaginary part associated to such a signal
since it is saturated and steady, and it has the shape of an almost
pure sinusoid after reaching saturation (phase noise is discussed
later on in this paper). The oscillation frequency curve dispersion
(red dots) still has a square root-like shape of the perturbation.
Figure \ref{fig:The-Puiseux-fractional-1} shows also another important
aspect, the flexibility in choosing the gain value in the nonlinear
circuit, i.e., different levels of mismatch between gain and loss,
using different values for the small-signal negative conductance $G_{1}=G_{e}(1+\delta)$
where $\delta=0,\;0.001\;$and$\;0.01$, represents the mismatch between
the loss and gain side (we recall that $G_{2}=G_{e}$). As shown in
Fig. \ref{fig:The-Puiseux-fractional-1}, even with $1\%$ mismatch
between gain and loss, the nonlinear circuit shows the same behavior
in the perturbation of the oscillation frequency, that is even matched
to the case with $\delta=0$. Thus, working in the unstable oscillation
configuration using nonlinearity in the coupled circuit gives us freedom
to tune the gain component's value and it works well even with some
mismatch between gain and loss. Note that the oscillation frequency
is highly sensitive to the capacitance perturbation on either side
of the circuit, either on the loss or gain side. Though not shown
explicitly, we have observed this feature theoretically, by calculating
the eigenfrequencies from $\mathrm{det}(\underline{\boldsymbol{\mathit{\mathrm{M}}}}-j\omega\underline{\boldsymbol{\mathrm{I}}})=0$
when varying $C_{1}$, and also verified the shifted resonance frequencies
using the prediction provided by the Puiseux series. Also, we have
observed in time-domain analyses with Keysight ADS circuit simulators
using nonlinear gain, that the shift of the oscillation frequency
is more sensitive to perturbation of $C_{1}$ than $C_{2}$. In this
paper, however, we only show the result from perturbing $C_{2}$ because
we want to investigate how a telemetric sensor works (i.e., the sensing
capacitance is on the passive part of the coupled resonators circuit).

\begin{figure}[tbh]
\begin{centering}
\includegraphics[width=3.2in]{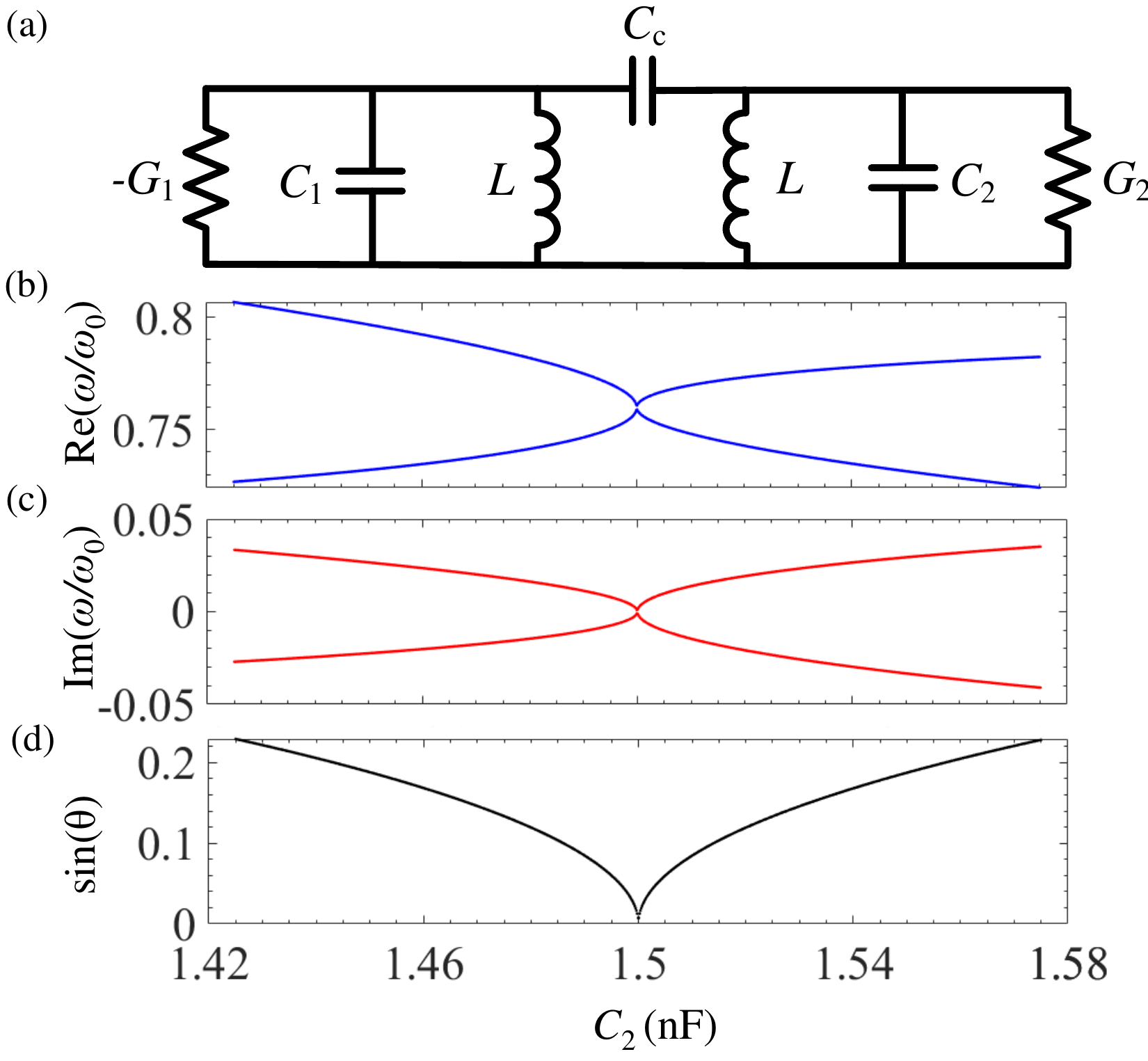}
\par\end{centering}
\centering{}\caption{\label{fig:Circuit_c}(a) Coupled resonators terminated with gain
$-G_{1}$ and loss $G_{2}$, with $G_{1}=G_{2}=G_{e}=9\:\mathrm{mS}$,
and $L=10\:\mathrm{\mu H}$, coupling capacitance $C_{c}=1.5\:\mathrm{nF}$,
capacitances $C_{1}=C_{2}=C_{e}=1.5\:\mathrm{nF}$. These parameters
lead to an EPD. The isolated (i.e., without coupling) resonance frequency
of each LC resonator is $\omega_{0}=1/\sqrt{LC_{e}}=25.8\times10^{6}\:\mathrm{s}^{-1}$.
The eigenfrequencies of the coupled circuit are calculated by solving
$\mathrm{det}(\underline{\boldsymbol{\mathit{\mathrm{M}}}}-j\omega\underline{\boldsymbol{\mathrm{I}}})=0$.
(b) Real and (c) imaginary parts of the angular eigenfrequencies normalized
by $\omega_{0}$, varying $C_{2}$ around the EPD value $C_{e}$.
(d) At the EPD, the coalescence parameter $\mathrm{sin}(\theta)$
vanishes, indicating that the two state vectors coalesce.}
\end{figure}

\section{Experimental demonstration of high sensitivity: case with coupling
capacitance\label{sec:Wired-Coupling}}

An analogous system with the properties highlighted in the previous
sections is made by the two resonators with balanced gain and loss
(PT-symmetry) coupled with a capacitor $C_{c}$ as shown in Fig. \ref{fig:Circuit_c}.
We discuss the condition to have an EPD and show the high sensitivity
theoretically and experimentally. First, we find the EPD condition
by writing down Kirchhoff's laws and using the Liouvillian formalism
using the system vector $\boldsymbol{\Psi}=[Q_{1},Q_{2},\dot{Q}_{1},\dot{Q}_{2}]^{\mathrm{T}}$,
where $Q_{n}$ is the capacitor charge on the gain side ($n=1$) and
the lossy side ($n=2$), and $\dot{Q}_{n}=dQ_{n}/dt$, leading to

\begin{equation}
\begin{array}{c}
\frac{d\boldsymbol{\Psi}}{dt}=\underline{\boldsymbol{\mathrm{M}}}\boldsymbol{\Psi}\\
\\
\underline{\boldsymbol{\mathrm{M}}}=\frac{1}{A}\left(\begin{array}{cccc}
0 & 0 & A & 0\\
0 & 0 & 0 & A\\
-\frac{B_{2}}{LC_{1}} & -\frac{C_{c}}{LC_{2}^{2}} & \frac{GB_{2}}{C_{1}} & -\frac{GC_{c}}{C_{2}^{2}}\\
-\frac{C_{c}}{LC_{1}^{2}} & -\frac{B_{1}}{LC_{2}} & \frac{GC_{c}}{C_{1}^{2}} & -\frac{GB_{1}}{C_{2}}
\end{array}\right)\\
\\
A=1+\frac{C_{c}}{C_{1}}+\frac{C_{c}}{C_{2}},\:\:B_{1}=1+\frac{C_{c}}{C_{1}},\:\:B_{2}=1+\frac{C_{c}}{C_{2}}.
\end{array}\label{eq:Liouvillian-2}
\end{equation}

\begin{figure}[t]
\begin{centering}
\includegraphics[width=3.5in]{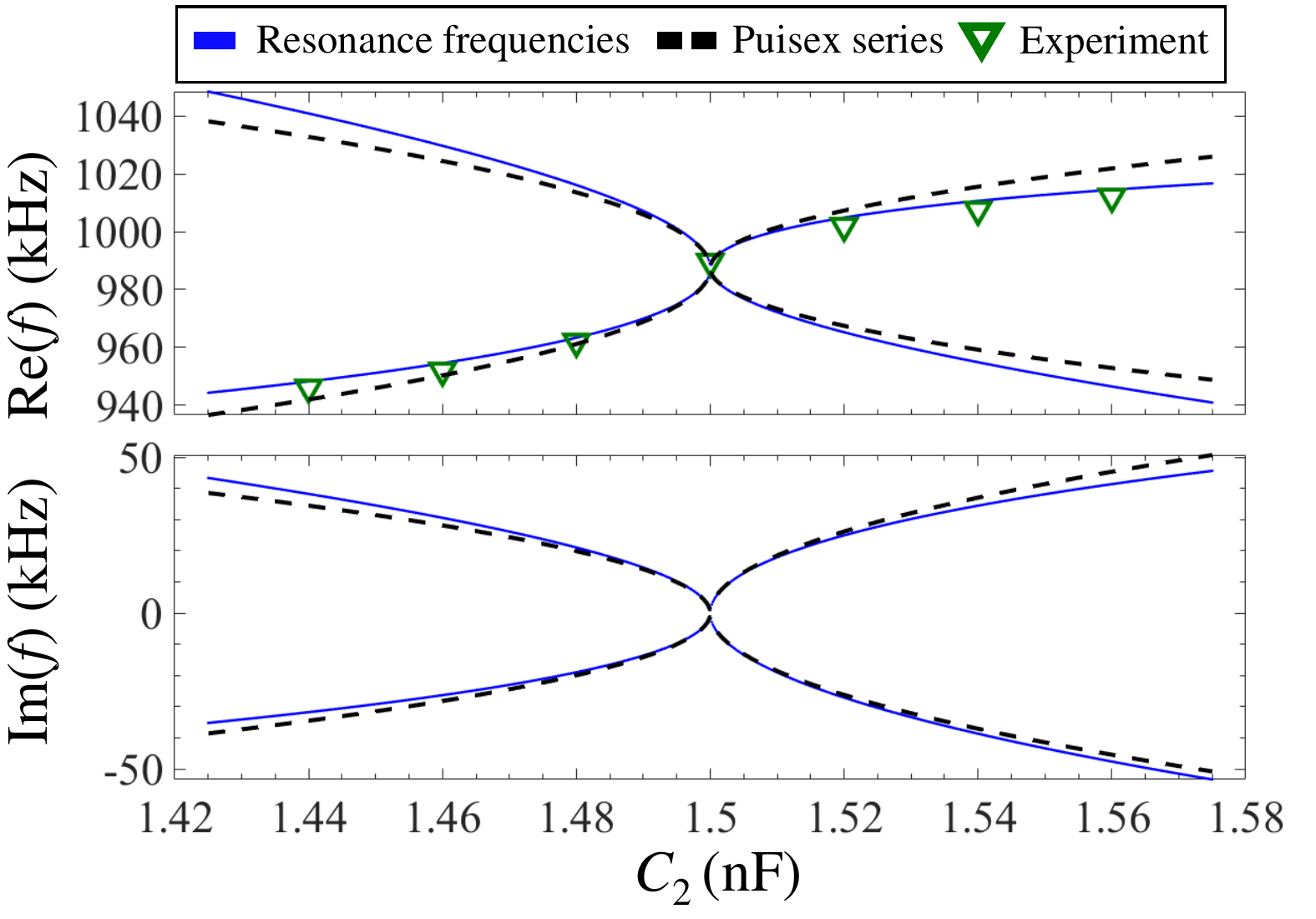}
\par\end{centering}
\caption{\label{fig:Experiment_freq}Experimental proof of exceptional sensitivity.
(a) Experimental and theoretical changes in the real part of the resonance
frequencies $f$ due to a positive and negative relative perturbation
$\Delta_{\mathrm{C_{2}}}$ applied to the capacitance $C_{2}$ as
$(1+\Delta_{\mathrm{C_{2}}})C_{e}$. Solid blue lines: eigenfrequencies
calculated by finding the zeros of the dispersion equation $\mathrm{det}(\underline{\boldsymbol{\mathit{\mathrm{M}}}}-j\omega\underline{\boldsymbol{\mathrm{I}}})=0$
using linear gain $G_{1}=G_{2}=G_{e}=9\:\mathrm{mS}$; dashed-black:
an estimate using the Puiseux fractional power expansion truncated
to its first order, using linear gain. Green triangles: oscillation
frequency measured experimentally (using nonlinear gain) after reaching
saturation for different values of $C_{2}$. The measured oscillation
frequency significantly departs from the EPD frequency $f_{e}=988.6\:\mathrm{kHz}$
even for a very small variation of the capacitance, approximately
following the fractional power expansion $f(\Delta_{\mathrm{C_{2}}})-f_{e}\propto\mathrm{Sgn}(\Delta_{\mathrm{C_{2}}})\sqrt{|\Delta_{\mathrm{C_{2}}}|}$.
Note that both positive and negative capacitance perturbations are
measured.}
\end{figure}

\begin{figure*}[!t]
\begin{centering}
\includegraphics[width=0.85\textwidth]{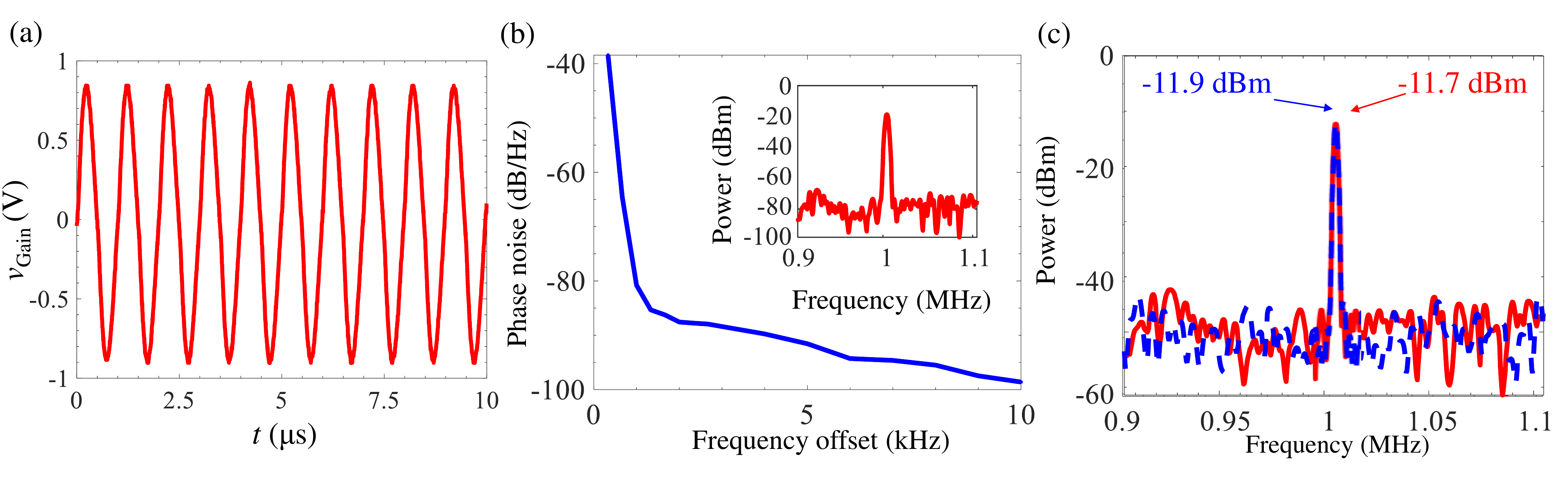}
\par\end{centering}
\caption{\label{fig:Time_phase noise}(a) Measured time-domain voltage signal
at the capacitor $C_{1}$ using an oscilloscope, when the system is
perturbed from EPD by $C_{\mathrm{2}}-C_{\mathrm{e}}=20\:\mathrm{pF},$
corresponding to a $\Delta_{\mathrm{C_{2}}}=0.013$. (b) Measured
wideband spectrum by Spectrum Analyzer (Rigol DSA832E) signal analyzer
as an inset with a fundamental frequency of oscillation of 1.002 MHz
(theoretical expectation based on $\mathrm{det}(\underline{\boldsymbol{\mathit{\mathrm{M}}}}-j\omega\underline{\boldsymbol{\mathrm{I}}})=0$
is at 1.004 MHz). Phase noise of the power spectrum is measured by
the Spectrum Analyzer at frequency offsets from a few Hz to 10 kHz.
The resolution bandwidth is set to 300 Hz, while video bandwidth is
set to 30 Hz to fully capture the spectrum. (c) Measured power spectrum
corresponding to a perturbation $\Delta_{\mathrm{C_{2}}}=0.013$ applied
to $C_{2}$, using two different gain values: the red curve is based
on gain of the EPD, and the blue curve is based on a gain that is
around 1\% larger than the EPD value.}
\end{figure*}

In this configuration, EPD occurs at $C_{1}=C_{2}=C_{c}=C_{e}=1.5\:\mathrm{nF}$,
linear gain and loss $G_{1}=G_{2}=G_{e}=9\:\mathrm{mS}$, $L=10\:\mathrm{\mu H}$.
Figures \eqref{fig:Circuit_c}(b) and (c) show the real and imaginary
parts of the eigenfrequencies when perturbing $C_{2}$, and Fig. \eqref{fig:Circuit_c}(d)
demonstrates the convergence of eigenvectors when $C_{2}=C_{e}$,
calculated by solving the dispersion equation $\mathrm{det}(\underline{\boldsymbol{\mathit{\mathrm{M}}}}-j\omega\underline{\boldsymbol{\mathrm{I}}})=0$
for $\omega$. The coalescence of two eigenvectors is observed by
defining the angle between them as in (\ref{eq:theta}), and this
indicates the coalescence of the two eigenmodes in their eigenvalues
and eigenvectors, and hence the occurrence of a second-order EPD.
It is seen from this plot that the system eigenfrequencies are coalescing
at a specific capacitance $C_{2}=C_{e}$ . The system is unstable
for any $C_{2}\neq C_{e}$ because of broken PT symmetry, since there
is always an eigenfrequency with $\mathrm{Im}(\omega)$<0. Moreover,
the bifurcation of the dispersion diagram at the EPD is in agreement
with the one provided by the Puiseux fractional power series expansion
truncated to its first order, represented by a dashed black line in
Fig. \eqref{fig:Experiment_freq}. The Puiseux series coefficient
is calculated as $\alpha_{1}=1.084\times10^{6}+j1.43\times10^{6}\:\mathrm{rad/s}$
by using Eq. (\ref{eq:alpha1}), assuming negative \textit{linear}
gain. The coefficient $\alpha_{1}$ is a complex number that implies
that the system always has two complex eigenfrequencies, for any $C_{2}$
value; that results in an unstable circuit, since one eigenfrequency
has $\mathrm{Im}(\omega)<0$, for any $C_{2}$ value.

In order to confirm the high sensitivity to a perturbation in the
proposed oscillator scheme based on nonlinear negative conductance
(nonlinear gain), the gain is now realized using an opamp (Analog
Devices Inc., model ADA4817) whose gain is tuned with a resistance
trimmer (Bourns Inc., model 3252W-1-501LF) to reach the proper small-signal
gain value of $-G_{1}=-9\:\mathrm{mS}$. Note that we assume that
the nonlinear gain is a bit larger (around 0.1 \%) than the loss on
the other side of the circuit to make the system slightly unstable.
All the other parameters are as in the previous example: a linear
conductance of $G_{2}=$ $9\:\mathrm{mS}$, capacitors of $C_{1}=C_{2}=C_{c}=1.5\:\mathrm{nF}$,
and inductors of $L=10\:\mathrm{\mu H}$ (Coilcraft, model MSS7348-103MEC).
This nonlinear circuit oscillates at the EPD frequency. The actual
experimental circuit differs from the ideal one using nonlinear gain
in a couple of points: First, extra losses are present in the reactive
components associated with their quality factor. The inductor has
the lowest quality factor in this circuit with an internal DC resistance
of $45\:\mathrm{m}\Omega$, from its datasheet, which is however small.
Second, electronic components have tolerances. To overcome some of
the imperfections in the experiment process, we use a capacitance
trimmer (Sprague-Goodman, model GMC40300) and a resistance trimer
in our printed circuit board (PCB) to tune the circuit to operate
at the EPD. Also, to have more tunability, a series of pin headers
are connected parallel to the loss side, where extra capacitors and
resistors could be connected in parallel, as mentioned in Appendix
\ref{subsec:Experiment}. The circuit is designed to work at the EPD
frequency of $f_{e}=988.6\:\mathrm{kHz}$, and indeed after tuning
the circuit, we experimentally obtain an experimental EPD frequency
at $f=989.6\:\mathrm{kHz}$ as shown in Fig. \ref{fig:Experiment_freq}
with a green triangle at $C_{2}=C_{e}$, very close to the designed
one. The oscillation frequency is obtained by taking the FFT of the
experimentally obtained time-domain voltage signal of the capacitor
$C_{1}$ using an oscilloscope (Agilent Technologies DSO-X 2024A)
after the signal reaches saturation for a time window of $10^{2}$
periods with $10^{6}$ points. The oscillation frequency is in agreement
with the result read by the spectrum analyzer (Rigol, model DSA832E).

We then perturb $C_{2}$ as $(1+\Delta_{\mathrm{C_{2}}})C_{e}$ where
$C_{e}$ satisfies the EPD condition, with small steps $\Delta_{\mathrm{C_{2}}}$
as explained in Appendix \ref{subsec:Experiment}. As shown in Fig.
\ref{fig:Experiment_freq}, the measured oscillation frequency dramatically
shifts away from the EPD frequency, following the square root of $\Delta_{\mathrm{C_{2}}}$
as theoretically predicted by Eq. (\ref{eq:Puisex}) for the linear
case. The experimental results (green triangles) in Fig. \ref{fig:Experiment_freq}
demonstrate that even for a small positive and negative perturbation
$C_{\mathrm{2}}-C_{\mathrm{e}}=\pm20\:\mathrm{pF},$ corresponding
to a $\Delta_{\mathrm{C_{2}}}=\pm0.013$, the oscillation frequency
significantly changes, which can be easily detected even in practical
noisy electronic systems. Figure \ref{fig:Time_phase noise}(a) shows
the experimental time-domain voltage signal of the capacitor $C_{1}$
with respect to the ground, when a relative perturbation $\Delta_{\mathrm{C_{2}}}=0.013$
is applied to $C_{2}$, measured by an oscilloscope. The spectrum\textquoteright s
frequency is now measured with a spectrum analyzer, and shown in \ref{fig:Time_phase noise}(b)
as an inset. The frequency of the spectrum matches the perturbed ($\Delta_{\mathrm{C_{2}}}=0.013$)
oscillation frequency, green triangle in Fig. \ref{fig:Experiment_freq},
obtained from the Fourier transform of the time domain experimental
data. These results confirm that the structure is oscillating at the
predicted perturbed resonance condition after saturation.

An essential feature of any oscillator is its ability to produce a
near-perfect periodic time-domain signal (pure sinusoidal wave), and
this feature is quantified in terms of phase noise, determined here
based on the measured power spectrum up to $10\:\mathrm{kHz}$ frequency
offset. The phase noise and power spectrum in Fig. \ref{fig:Time_phase noise}(b)
demonstrate that electronic noise (which is significant in opamp)
and thermal noise in the proposed highly sensitive oscillator scheme
does not discredit the potential of this circuit to exhibit measurable
high sensitivity to perturbations. Indeed, the low phase noise of
$-80.8\:\mathrm{dB/Hz}$ at $1\:\mathrm{kHz}$ offset from the oscillation
frequency shows that the frequency shifts observed in Fig. \ref{fig:Experiment_freq}
are well measurable. Note that this result is intrinsic in the nonlinear
saturation regime proper of an oscillator. The resonance oscillation
peaks have a very narrow bandwidth (linewidth), which makes the oscillation
frequency shifts very distinguishable and easily readable. In this
oscillator-sensor system, we also have some freedom in choosing the
small-signal gain value because the dynamics are also determined by
the saturation arising from the nonlinear gain behavior. For example,
in the experiment, we have verified that circuit has the same oscillation
frequency when using an unbalanced small-signal gain $1\%$ larger
than the balanced loss value. Figure \ref{fig:Time_phase noise}(c)
shows two measured frequency spectra corresponding to a relative perturbation
$\Delta_{\mathrm{C_{2}}}=0.013$ applied to $C_{2}$, using two different
gain values. The spectrum has been measured using a resolution bandwidth
of 300 Hz, while the video bandwidth is set to 30 Hz to fully capture
the spectrum. The red curve is for case with gain around $1\%$ bigger
than the balanced loss whereas the blue curve is for the case where
gain and loss are balanced. These two frequency responses show the
same oscillation frequency but the power spectrum peak has a very
small difference, 0.2 dBm higher for the case with $1\%$ larger gain
shown in Fig.\ref{fig:Time_phase noise}(c). This important feature
that helps us design the circuit without a very accurate balance between
gain and loss, i.e., oscillator-sensors can be realized without satisfying
exactly PT symmetry also when the sensing perturbation is not applied.
As mentioned earlier, the nonlinear oscillator with broken PT symmetry
exhibits the very important feature that the oscillation frequency
shifts are both positive and negative, depending on the sign of the
perturbations $\Delta_{\mathrm{C_{2}}}$, hence allowing sensing positive
and negative values of $\Delta_{\mathrm{C_{2}}}$.

\section{Conclusions}

We demonstrated that two coupled LC resonators terminated with nonlinear
gain, with almost balanced loss and small-signal gain, working near
an EPD, make an oscillator whose oscillation frequency is very sensitive
to perturbations. The nonlinear behavior of the active component is
essential for the three important features observed by simulations
and experimentally: (i) the oscillation frequency is very sensitive
to perturbations, and both positive and negative perturbations of
a capacitor are measured leading to very high sensitivity based on
shifted oscillation frequency that approximately follows the square-root
law, proper of EPD systems; (ii) the measured spectrum has very low
phase noise allowing clean measurements of the shifted oscillation
frequencies. (iii) It is not necessary to have a perfect gain/loss
balance, i.e., we have shown that slightly broken gain/loss balance
leads to the same results as for the case of perfectly balanced gain
and loss.
\begin{figure}[t]
\centering{}\includegraphics[width=1.5in]{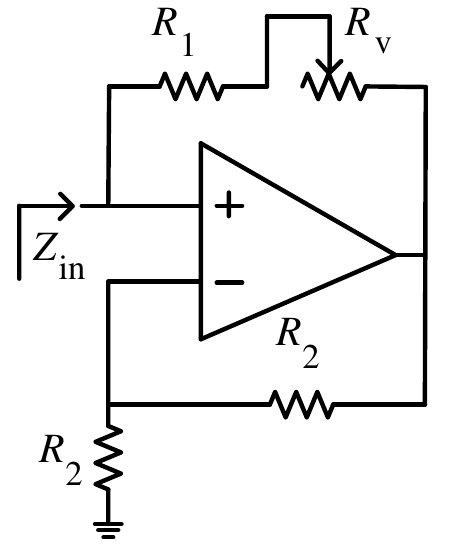}\caption{\label{fig:NegativeR}Negative resistance converter circuit implementation
by using an opamp.}
\end{figure}

Note that none of the features above are available in current PT-symmetry
circuits in the literature \cite{Schindler2011Experimental,Chen2018Generalized}:
Indeed, only one sign of the perturbation is measurable with the PT-symmetry
circuits published so far, since the other sign leads to the circuit
instability. Furthermore, to make a single sign perturbation measurement,
in the literature, e.g., \cite{Chen2018Generalized}, the capacitor
$C_{1}$ on the gain side has been tuned using a varactor to reach
the value of the perturbed capacitor ($C_{2}$) on the reading side
in order to rebuild the PT symmetry (but in a sensor operation it
is not possible to know a priori the value that has to be measured);
furthermore, to work at or very close to an EPD, using linear gain,
the gain has to be set equal to the loss (balanced gain/loss condition).

The oscillation frequency shift follows the square root-like behavior
predicted by the Puiseux series expansion, as expected for EPD-based
systems. We show the performance of the oscillator-sensor scheme based
on two configurations: wireless coupling with a mutual inductor, and
wired coupling by a capacitor. The latter oscillator scheme has been
fabricated and tested. We have analyzed how the nonlinearity in the
gain element makes the circuit unstable and oscillate after reaching
saturation. The oscillator's characteristics have been determined
in terms of transient behavior and sensitivity to perturbations due
to either capacitance or resistance change in the system. The experimental
verification provided results in very good agreement with theoretical
expectations. The measured high sensitivity of the oscillator sensor
to perturbations can be used as a practical solution for enhancing
sensitivity, also because the measured shifted frequencies are well
visible with respect to underlying noise. The proposed EPD-based oscillator-sensor
can be used in many automotive, medical, and industrial applications
where detections of small variations of physical, chemical, or biological
variations need to be detected.
\begin{figure}[t]
\centering{}\includegraphics[width=3.5in]{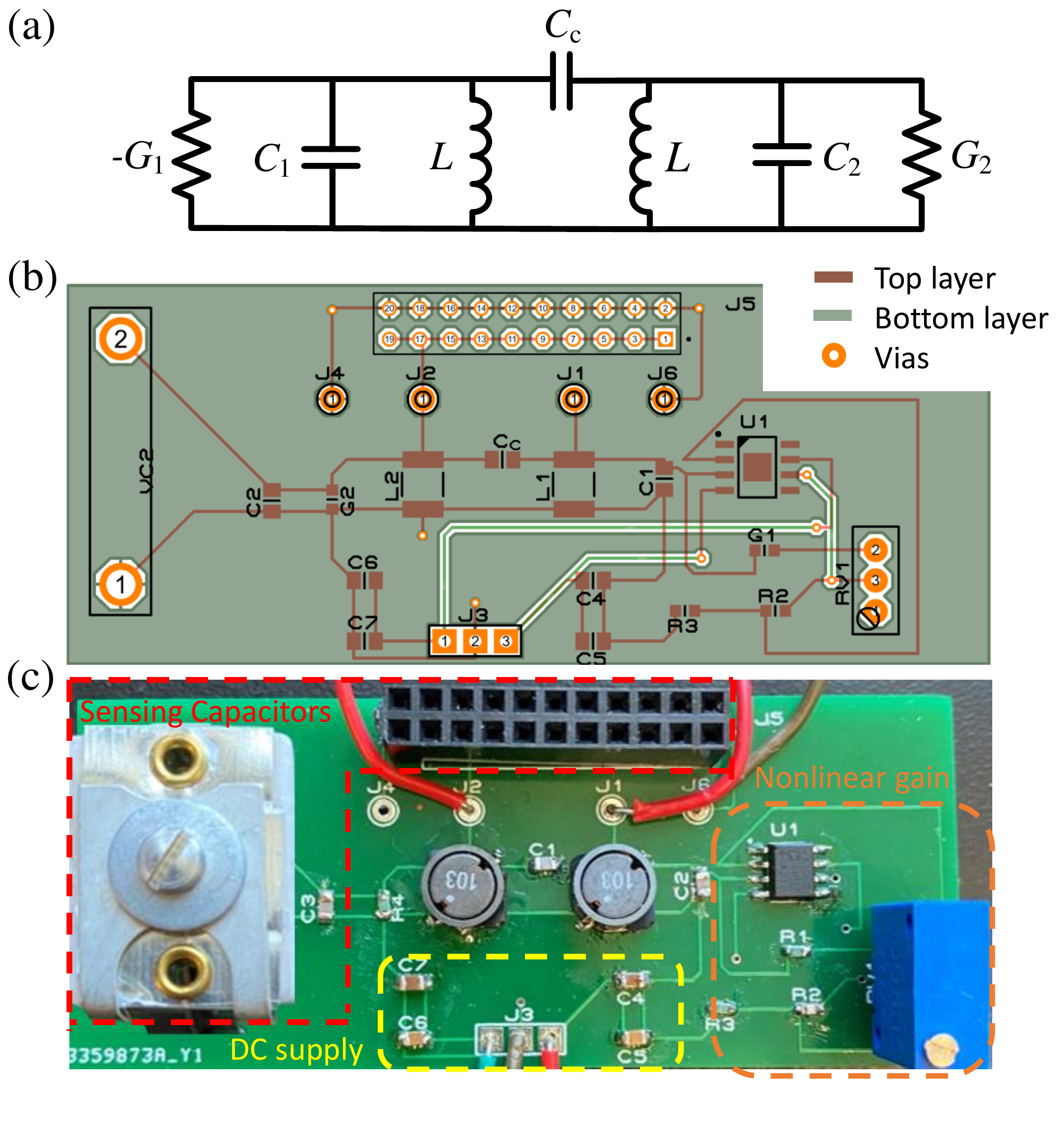}\caption{\label{fig:Implemented-Circuit}(a) Schematic of the two LC resonators
coupled by $C_{\mathrm{c}}=1.5\:\mathrm{nF}$ with inductor $L_{1}=L_{2}=10\:\mathrm{\lyxmathsym{\textmu}H}$,
the opamp $\mathrm{U}_{1}$ (Analog Devices Inc., model ADA4817),
the variable resistance $\mathrm{RV_{1}}$ (Bourns Inc., model 3252W-1-501LF)
and variable capacitance $V_{C2}$ (Sprague-Goodman, model GMC40300),
biasing capacitors $C_{4}=C_{6}=0.1\:\mathrm{\mu F}$ $C_{5}=C_{6}=10\:\mathrm{\mu F}$
as datasheet suggested. (b) PCB layout of the assembled circuit where
the top layer traces are red, the ground plane and bottom traces are
green, and the connecting vias are orange. In this design, Via $\mathrm{J1}$
is a probe point for the capacitor voltage, whereas Vias $\mathrm{J6}$
and $\mathrm{J4}$ are test points connected to the ground plane and
are used to connect the ground of the measurement equipment to the
ground of the circuit. All the ground nodes are connected to the bottom
green layer.}
\end{figure}

\section*{Acknowlegment}

This material is based upon work supported by the USA National Science
Foundation under Award NSF ECCS-1711975.

\appendices{}

\section{Negative resistance}

Several different approaches provide negative nonlinear conductance
needed for proposed circuits. In this subsection, we show the circuit
in Fig. \ref{fig:NegativeR} that utilizes opamp to achieve negative
impedance. The converter circuit converts the impedance as $Z_{in}=-R_{1}$
while we design the circuit to work at the EPD point by choosing $R_{1}=1/G_{e}$.
In the experiment, we used $R_{1}=100\:\Omega$, and $R_{2}=2\:\mathrm{k}\Omega$
to achieve the EPD value. We tuned the negative resistance with resistor
trimmer $R_{\mathrm{v}}$ to reach the EPD value $G_{e}=9\:\mathrm{ms}$.

\section{Implementation of the nonlinear coupled oscillator\label{subsec:Experiment}}

We investigate resonances and their degeneracy in the two LC resonators
coupled by a capacitor as in Fig. \ref{fig:Implemented-Circuit}(a),
whereas Figs. \ref{fig:Implemented-Circuit}(b) and (c) illustrate
the PCB layout and assembled circuit. In the fabricated circuit, the
sensing capacitance is shown in the red dashed box, the nonlinear
gain is in the orange dashed box, and the DC supply is in the yellow
dashed box. Inductors have values $L_{1}=L_{2}=10\:\mathrm{\lyxmathsym{\textmu}H}$,
the loss value is set to $G_{2}=9\:\mathrm{mS}$ with a linear resistor,
the capacitor on the gain side and the coupling capacitor are $C_{1}=C_{c}=1.5\:\mathrm{nF}$.
The gain element is designed with an opamp (Analog Devices Inc., model
ADA4817), where the desired value of gain is achieved with a variable
resistor RV1. In the experiment, we setup the nonlinear gain to be
a bit larger (around 0.1\%) than the balanced gain by tuning the RV1
to make the system slightly unstable.To tune and find the exact value
of the capacitance that leads to an EPD $\left(C_{2}=C_{e}\right)$,
a variable capacitor (Sprague-Goodman, model GMC40300) and a series
of pin headers, where extra capacitors could be connected in parallel
to $C_{2}$, are provided. By adding small and known capacitors values
on the load side, we tuned the capacitance $C_{2}$ to bring the circuit
very close to the EPD and observe the EPD oscillation frequency $f=f_{e}$.

To show the square root-like behavior of the oscillator's frequency
due to perturbations in Fig. \ref{fig:Experiment_freq} and \ref{fig:Circuit_c},
we perturbed the capacitor $C_{2}$ with pairs of extra 10 pF capacitors
to make 20 pF steps, connected in parallel to $C_{2}$, using the
pin headers shown in Fig. \ref{fig:Implemented-Circuit}. After each
perturbation, the oscillation frequency is measured with an oscilloscope
and also with a spectrum analyzer (for comparison and verification
purposes), as discussed in Section \ref{subsec:Experiment}, and shown
in Fig. \ref{fig:Experiment_freq} with green triangles. Moreover,
for the perturbed circuit, considering $\Delta_{\mathrm{C_{2}}}=0.013$
applied to $C_{2}$ (any perturbed point can be chosen), we changed
the variable resistor RV1 to study oscillation frequency variation
for different unbalanced nonlinear gains. The goal was to show that
the circuit using a bit unbalancednonlinear gain still has the same
oscillation frequency. Indeed, by trimming the RV1, we verified the
same oscillation frequency for roughly $1\%$ unbalanced gain and
loss, as shown in Fig. \ref{fig:Time_phase noise}(c). Note that on
the PCB, the ground plane (on the bottom layer) is designed to connect
all the ground of the measurement equipments and DC supply to the
circuit's ground.


\end{document}